\documentclass[aps,twocolumn,showpacs,preprintnumbers,nofootinbib,prl,superscriptaddress,10pt]{revtex4-1}

\usepackage[utf8]{inputenc}
\usepackage{graphicx}
\usepackage{amsmath,amssymb}
\usepackage{amsfonts}
\usepackage{physics}
\usepackage{xspace}
\usepackage{bm}
\usepackage{mathrsfs}
\usepackage[dvipsnames]{xcolor}
\usepackage[unicode]{hyperref}
\hypersetup{colorlinks=true, citecolor=MidnightBlue, linkcolor=CornflowerBlue, urlcolor=CornflowerBlue, linktocpage=true}
\usepackage[all]{hypcap}
\usepackage{multirow}
\usepackage[format=plain,justification=RaggedRight,labelsep=endash,font=small,labelfont=sc]{caption}
\usepackage[export]{adjustbox}

\definecolor {darkgreen}{rgb}{0.2,0.7,0.2}

\newcommand{\eq}{\begin{equation}}
\newcommand{\be}{\begin{equation}}
\newcommand{\eeq}{\end{equation}}
\newcommand{\ee}{\end{equation}}

\newcommand*\DAlembert{\mathop{}\!\mathbin\Box}
\newcommand{\note}[1]{\text{\scshape\tiny{#1}}}

\begin{document}

\title{Spin-induced black hole spontaneous scalarization}
 \author{Alexandru Dima}
 \affiliation{SISSA, Via Bonomea 265, 34136 Trieste, Italy and INFN Sezione di Trieste}
 \affiliation{IFPU - Institute for Fundamental Physics of the Universe, Via Beirut 2, 34014 Trieste, Italy}
 \author{Enrico Barausse}
 \affiliation{SISSA, Via Bonomea 265, 34136 Trieste, Italy and INFN Sezione di Trieste}
 \affiliation{IFPU - Institute for Fundamental Physics of the Universe, Via Beirut 2, 34014 Trieste, Italy}
 \affiliation{Institut d'Astrophysique de Paris, CNRS \& Sorbonne
  Universit\'es, UMR 7095, 98 bis bd Arago, 75014 Paris, France}
 \author{Nicola Franchini}
 \affiliation{SISSA, Via Bonomea 265, 34136 Trieste, Italy and INFN Sezione di Trieste}
 \affiliation{IFPU - Institute for Fundamental Physics of the Universe, Via Beirut 2, 34014 Trieste, Italy}
 \author{Thomas P. Sotiriou}
 \affiliation{School of Mathematical Sciences \& School of Physics and Astronomy, University of Nottingham, University Park, Nottingham, NG7 2RD, UK}

\begin{abstract}
We study scalar fields in a black hole background and show that, when the scalar is suitably coupled to curvature, rapid rotation can induce a tachyonic instability. This instability, which is the hallmark of spontaneous scalarization in the linearized regime, is expected to be quenched by nonlinearities and endow the black hole with scalar hair. Hence, our results demonstrate the existence of a broad class of theories that share the same stationary black hole solutions with general relativity at low spins, but which exhibit black hole hair at sufficiently high spins ($a/M\gtrsim 0.5$). This result has clear implications for tests of general relativity and the nature of black holes with gravitational and electromagnetic observations.
\end{abstract}

\date{\today \hspace{0.2truecm}}

\maketitle
\flushbottom

\noindent\textbf{\textit{Introduction:}}
Direct and indirect detections leave little doubt that black holes (BH)  exist in nature~\cite{Webster:1972bsw,1999ApJ...524..816R,Schodel:2002vg,Reid_2003,Gillessen:2008qv,Abbott:2016blz,2018A&A...618L..10G,Akiyama:2019cqa}. In general relativity (GR) the mass and the spin of an astrophysical BH  fully determine its properties. An electric charge is also technically allowed, but is expected to be paltry for
astrophysical BHs, see e.g.~\cite{Barausse:2014tra}.  Any other quantity, \textit{hair} in jargon, is not necessary according to no-hair theorems~\cite{Israel:1967wq,Carter:1971zc,Robinson:1975bv}. Future gravitational wave detectors will finally allow us to confront theorems and observations with unprecedented precision~\cite{Berti:2016lat,Barausse:2016eii,Toubiana:2020vtf}, improving upon current observations, which are perfectly compatible with hairless BHs~\cite{LIGOScientific:2019fpa,Isi:2019aib}.

It is tempting to interpret an absence of BH hair as a vindication of GR minimally coupled to the Standard Model. However, new fundamental fields can be more elusive. It is illustrative to consider scalar fields: no-hair theorems exist for stationary BHs in scalar-tensor theories~\cite{Hawking:1972qk,Sotiriou:2011dz}, and static, spherically symmetric and slowly rotating BHs in shift-symmetric generalized (Horndeski) scalar-tensor theories~\cite{Hui:2012qt,Sotiriou:2013qea}.\footnote{No-hair theorems also
exist for stars in shift-symmetric scalar tensor theories~\cite{starsNohair1,starsNohair2,Lehebel:2017fag,Yagi:2015oca}.}
In fact, it turns out that there is a single coupling term in the Horndeski class that gives rise to hair: a linear coupling between the scalar and the Gauss-Bonnet (GB) invariant~\cite{Sotiriou:2013qea,Sotiriou:2014pfa}, given by
\begin{equation}\label{eq:GBinv}
\mathcal{G}=R^{\mu\nu\rho\sigma}R_{\mu\nu\rho\sigma}-4~R^{\mu\nu}R_{\mu\nu}+R^2.
\end{equation}
Considering that the Horndeski class contains all actions for a massless scalar nonminimally coupled to gravity that yield second order equations upon variation, absence of hair actually seems to be the norm rather than the exception for scalar fields. Indeed, known hairy BH solutions circumvent theorems by evading one or more of their assumptions, see {\em e.g.}~\cite{Sotiriou:2013qea,Babichev:2013cya,Cardoso:2013fwa,Cardoso:2013opa,Herdeiro:2014goa,Antoniou:2017acq,Sotiriou:2015pka}.

A further  complication in attempting to detect new fields through BH hair is the possibility that, even within the context of the same theory, only certain BHs might actually exhibit it.
This was realized only recently, as the first models of BH {\em scalarization} appeared in the literature~\cite{Silva:2017uqg,Doneva:2017bvd}. For concreteness, consider the action
\begin{equation}\label{eq:gbaction}
S=\frac{1}{2}\int d^4x \sqrt{-g}\left(R-\frac{1}{2}\nabla_{\mu}\phi\nabla^{\mu}\phi+
f(\phi)\mathcal{G}\right),
\end{equation}
where $f$ is some function of $\phi$,
and where we have also set (as in the rest of this paper) $8\pi G = c =1$.
Varying the action with respect to $\phi$ yields
\begin{equation}\label{eq:ScalarEq}
\DAlembert \phi = - f'(\phi)\mathcal{G},
\end{equation}
where $f'(\phi)\equiv df/d\phi$. Assume that $f'(\phi_0)=0$, for some constant $\phi_0$. Then solutions with $\phi=\phi_0$ are admissible and they are also solutions of GR.
A no-hair theorem~\cite{Silva:2017uqg} ensures that they are unique if they are stationary, provided that $f''(\phi)\mathcal{G} < 0$.

The fact that GR BHs are stationary solutions to this theory is not sufficient to conclude that there are no observable deviations from GR, as  the  perturbations
over these solutions  do not generally obey the GR field equations~\cite{Barausse:2008xv}. These perturbations
may even grow unstable, thus rendering the GR  solutions irrelevant.
 Indeed, one can think of $-f''\mathcal{G}$ as the (square of the) mass of the scalar perturbation on a fixed background.
Hence, the condition above ensures that this effective (squared) mass is positive. If the condition is violated and the effective (squared) mass becomes sufficiently negative, the GR solutions suffer a tachyonic instability and the scalar develops a nontrivial profile.

A similar scalarization effect  was shown to occur for neutron stars in a different class of scalar-tensor theories more that 25 years ago~\cite{Damour:1993hw},
and is triggered when the star compactness reaches a critical threshold. Related ``dynamical'' scalarization effects~\cite{ST1,ST2,ST3,ST4}
are present in the same theories for neutron star binaries, whenever their separation
is sufficiently small (or the binary's ``compactness'' sufficiently large). However, in the class of theories considered
in~\cite{Damour:1993hw,ST1,ST2,ST3,ST4},  scalarization is not present without matter,  and BHs are vacuum solutions.\footnote{Black holes can scalarize if they have matter in their vicinity~\cite{Cardoso:2013fwa,Cardoso:2013opa}, but the densities necessary to
obtain a measurable effect are probably astrophysically unrealistic.}

Black hole scalarization is fairly well understood. It starts as a linear tachyonic instability and, as such, its onset is controlled only by terms that contribute to linear perturbations around GR solutions. In this sense,  action~\eqref{eq:gbaction} with $f(\phi)=\eta\phi^2/2$ is sufficient to study the onset of scalarization~\cite{Silva:2017uqg,Andreou:2019ikc}. As the instability develops and the scalar grows, nonlinear terms become increasingly important and eventually quench the instability. Hence, the endpoint and properties of the scalarized solutions are actually controlled by the nonlinear interactions of the scalar~\cite{Silva:2018qhn,Macedo:2019sem}. A characteristic example is that in models with different nonlinear interactions, scalarized solutions have different stability properties~\cite{Blazquez-Salcedo:2018jnn,Silva:2018qhn}.

Here we will focus exclusively on the onset of scalarization, so we will restrict our attention to
quadratic scalar GB (qsGB) gravity, i.e.  $f(\phi)=\eta\phi^2/2$ (without loss of generality \cite{Andreou:2019ikc}). The effective (squared) mass of the scalar on a fixed background is then
\begin{equation}\label{meff}
	\mu^2_\note{eff} = -\eta \mathcal{G}.
\end{equation}
For the Schwarzschild solution, one has $\mathcal{G}=48 M^2/r^6$, which is always positive and decreasing with $r$,  and which yields
the horizon value $\mathcal{G}(r=2M)=3/(4M^4)$. Hence,  a tachyonic instability only occurs for $\eta>0$, and the instability is expected to be more violent for smaller masses.\footnote{Note that in curved spacetimes $\mu^2_\note{eff}$ can be somewhat negative without necessarily developing a tachyonic instability.} This is why the focus in the literature so far has been on $\eta>0$ (or the equivalent condition in more complicated models). However, for a Kerr BH of mass $M$ and spin parameter $a$ in Boyer-Lindquist $(t,r,\theta,\varphi)$ one has
\begin{equation}\label{eq:gbterm}
\mathcal{G}_\note{Kerr}=\frac{48 M^2}{(r^2+\chi^2)^6}\left(r^6-15r^4\chi^2+15r^2\chi^4-\chi^6\right)
\end{equation}
where, for brevity, $\chi\equiv a\cos\theta$. Clearly, $\mathcal{G}_\note{Kerr}$ is not monotonic, and can even become negative close to the horizon. This explains the results of~\cite{Cunha:2019dwb,Collodel:2019kkx}, where it was shown that rotation suppresses scalarization for $\eta>0$.

In this Letter we focus on $\eta<0$, which yields a real effective mass $\mu_\note{eff}$ for low BH spins, but  which  can  yield an imaginary $\mu_\note{eff}$ for high spins. We investigate the
behavior of linear scalar perturbations to the GR solution by evolving Eq.~\eqref{eq:ScalarEq} on a Kerr background, with
the goal of assessing for what BH spins and couplings $\eta$ the perturbations become unstable.
Indeed, at least two possible instability mechanisms may be at play in Eq.~\eqref{eq:ScalarEq}. The first is the tachyonic instability associated to spontaneous scalarization, mentioned above. The second could be a superradiant instability, which is known to exist at high spins for constant real masses~\cite{Damour:1976,Zouros:1979iw,Detweiler:1980uk,Dolan:2007mj}, and potentially also for non-constant effective masses~\cite{Dima:2020rzg} such as the one of Eq.~\eqref{meff}. Superradiance occurs when bosonic waves
with non-vanishing angular momentum are amplified when scattered by a spinning BH, at the expense
of the rotational energy of the BH, which as a result spins down. For massive bosons, superradiant scattering
can develop into an instability because the field is confined near the BH by its own mass.

 It should be stressed that, in principle, both instabilites could be present. However, they have distinct features (timescales,
the angular momenta involved, dependence on the BH spin). We show below that the tachyonic instability is by far the dominant effect for  $\eta<0$. More broadly, our results strongly suggest that there exist theories in which scalarization occurs only for rapidly rotating BHs.

\noindent\textbf{\textit{Methodology:}}
For $f(\phi)=0$ and over a Kerr background, Eq.~\eqref{eq:ScalarEq} separates into ordinary differential equations when $\phi$ is decomposed onto
a basis of spheroidal harmonics. However, the choice $f(\phi)=\eta \phi^2/2$ yields an intrinsically non-separable equation.
We therefore resort to a time-domain numerical integration of this equation, by using techniques akin to
those presented in \cite{Dima:2020rzg,Dolan:2012yt}, to which we refer for more details.

In brief, the idea is to project Eq.~\eqref{eq:ScalarEq}  onto a basis of \textit{spherical}\footnote{There is no advantage in using spheroidal harmonics, for
which analytic expressions are unavailable, as they do not lead to a separable equation anyway.}
harmonics $\mathbf{Y}_{lm}$, which yields 1+1 evolutions equations (in $t$ and $r$) for
the components of the scalar field,
\begin{equation}\label{eq:decomp}
\psi_{lm}(t,r)\equiv \int \mathbf{Y}^*_{lm}(r\phi)d\Omega
\end{equation}
These equations are coupled and given explicitly by
\begin{align}\label{eq:perteq}
&\left[(r^2+a^2)^2-a^2\Delta(1- c^m_{ll})\right]\ddot{\psi}_{l}+a^2\Delta(c_{l,l+2}^m\ddot{\psi}_{l+2}+\nonumber\\
&+c_{l,l-2}^m\ddot{\psi}_{l-2})+4iamMr\dot{\psi}_l \nonumber\\
&-(r^2+a^2)^2\psi''_l-\left(2iam(r^2+a^2)-2a^2\frac{\Delta}{r}\right)\psi'_l\nonumber\\
&+\Delta \left[ l(l+1)+\frac{2M}{r}-\frac{2a^2}{r^2}+\frac{2iam}{r} \right]\psi_l\nonumber\\
&+\Delta\sum_{j}\langle l m | \mu^2_\note{eff} (r^2+\chi^2) | j m \rangle\psi_j=0 \,,\\
&\Delta\equiv r^2-2 M r +a^2	\,,\\
\label{eq:coscoup}
&c^m_{jl}\equiv \langle l m | \cos^2\!\theta | j m\rangle \nonumber\\&= \frac{\delta_{lj}}{3}+\frac{2}{3}\sqrt{\frac{2j+1}{2l+1}}\langle j,2,m,0|l,m\rangle \cdot \langle j,2,0,0|l,0\rangle\,,
\end{align}
where  $\langle j_1,j_2,m_1,m_2|j_3,m_3\rangle$ are the Clebsch-Gordan coefficients~\cite{AngMom}.
Note that the evolution of modes of different $m$
decouples because of the axisymmetry of the problem. Moreover, because of
reflection symmetry with respect to the origin,
even-$l$ and odd-$l$ modes also decouple: the evolution of
a mode $(l,m)$ is coupled to that of all the modes $(l+2k,m)$, with $k=1,2,3,\ldots$.

To numerically evolve the system \eqref{eq:perteq}, we discretize  the spatial grid and use a method of lines. By integrating in time using a fourth order explicit Runge-Kutta time-step inside  the computational grid (as done e.g in~\cite{Dima:2020rzg}), it becomes apparent that the equations
are stiff for large $\eta$, and that the numerical integration becomes unstable. To overcome this problem, we have used an Implicit-Explicit (IMEX) Runge-Kutta solver with adaptive time step, namely the IMEX-SSP3(3,3,2) and IMEX-SSP(4,3,3) schemes of~\cite{Palenzuela:2008sf}.
Note that implicit methods~\cite{PareschiRusso}, while effective at dealing with stiff  problems, are typically less accurate
and more computationally expensive. However, implicit-explicit algorithms, by employing explicit steps for the non-stiff terms and implicit steps only for the stiff ones, can tackle stiff problems with limited computational overhead.
We successfully compared our code to results  from both frequency-domain techniques~\cite{Berti:2009kk} and similar time-domain codes~\cite{Dolan:2012yt}. Our implementation was also tested by analysing the convergence of the results (and their overall robustness) vs time-step and spatial-grid resolution.


\noindent\textbf{\textit{Results:}}
To investigate the possible presence of an instability, we evolve the scalar field by integrating the system given by Eq.~\eqref{eq:perteq}, with $l$ ranging from 0 to $l_{\rm max}=30$ and $|m|\leq l$, and with Gaussian initial conditions for each mode $\psi_{lm}$. The results are
robust against the choice of the cutoff $l_{\rm max}$  -- as long as that is sufficiently large -- and initial conditions, which only affect the early transient evolution of the scalar and not the unstable growth phase, if present. We consider BH spins $a/M \sim 0.5$ -- $0.999$ and qsGB coupling $\abs{\eta}/ M^2 \sim 0.1$ -- $10^{5}$.

\begin{figure}[h!t]
	\includegraphics[scale=0.45]{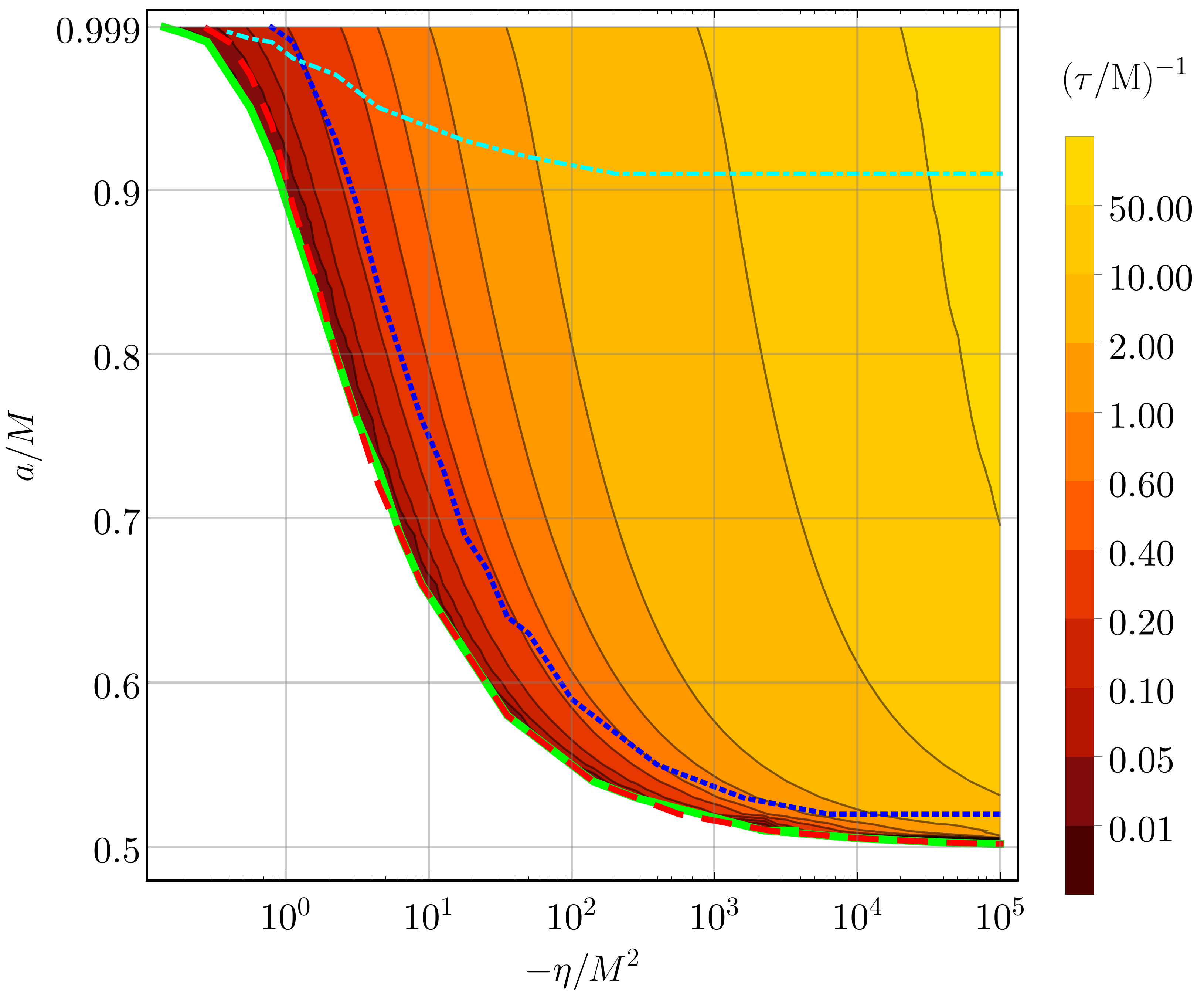}
	\caption{Instability timescale $\tau$ (color code) for the reconstructed field as a function of spin and GB coupling. The instability threshold
for the total reconstructed field is shown by the solid green line, while
the threshold when the $m=0$ modes are excluded is shown by a blue dotted line.
The red dashed line corresponds the instability threshold for the $m=0$ odd modes,
while the  dot-dashed cyan line marks the instability threshold for the spherical mode $l=m=0$ (see text for details). Note
that all shown values of $\eta$ are unconstrained by different observables (c.f. discussion in the conclusions).}	\label{fig:tacReconstructed}
\end{figure}

From the simulations showing an exponential scalar growth, we extract the instability timescale $\tau$ of the reconstructed field $|\phi|=\left(\sum_{lm} |\psi_{lm}|^2 \right)^{1/2}\propto \exp(t/\tau)$ by fitting the time evolution of the scalar's amplitude after the initial transient. The contours in Fig.~\ref{fig:tacReconstructed} show $\tau^{-1}$ as a function of $a/M$ and $\abs{\eta} /M^2$. The instability becomes stronger as either the spin or the coupling increases. Moreover, there is a
 minimum spin $a_{\rm min}$ below which the instability disappears. For  $|\eta|\to\infty$,
it appears that $a_{\rm min}/M\to 0.5$ (up to percent level numerical errors).
The solid green line denotes the combinations of parameters for which the instability disappears (i.e. $\tau\to\infty$). With the blue dotted line we show the same
marginal instability curve for the reconstructed field, but excluding the $m=0$ modes.
As can be seen, when the latter are excluded the parameter
space region yielding an instability shrinks, i.e.
the main contribution to the instability comes from the $m=0$ modes.
As a further test of this conclusion, we also computed the marginal instability curve
for the $m=0$ modes alone, and that does indeed match
the solid green line in Fig.~\ref{fig:tacReconstructed}.

Even and odd parity modes (i.e. modes with even and odd $l$) automatically decouple in Eq.~\eqref{eq:perteq}.
In the $m=0$ sector, which dominates the instability shown
in Fig.~\ref{fig:tacReconstructed},
the odd and even modes give roughly comparable contributions.
We have verified this by considering the marginal instability
curves for the odd and even $m=0$ modes separately,
which are both very close to the
solid green line of  Fig.~\ref{fig:tacReconstructed}.
As an example, the red dashed line in  Fig.~\ref{fig:tacReconstructed}
represents the marginal instability curve for
the $m=0$ odd modes.

Indeed, odd modes seem to have only marginally shorter instability times (by $\sim 1-2\%$) than even ones for high spins and large couplings.
Conversely,  in the region $\abs{\eta}<1$, $a/M>0.9$ the even modes
are slightly more unstable, as can be seen
from the somewhat increased distance between the red dashed and solid green line curves.

Next we consider if some individual angular mode $l,m$ gives the dominant contribution to the instability.
To answer this question, we have to override the non-separability of the problem. To this end,
we have forcefully decoupled each $l$-mode  in Eq.~\eqref{eq:perteq} , suppressing ``by hand'' all the couplings between angular modes (i.e. $\langle l m | \mu^2_\note{eff} (r^2+\chi^2) | j m \rangle$ with $l\neq j$) generated by the GB invariant; we have only kept active the contributions to the effective mass of the single $l$-mode. We have then let the system evolve, selecting Gaussian initial data for the chosen mode only.
By this technique, we have isolated, for instance, the
instability parameter space for the spherical mode $l=m=0$,
whose marginal instability curve is shown in Fig.~\ref{fig:tacReconstructed} by a cyan dot-dashed line.
However, we could not find any single $l,m$ mode for which the marginal instability curve
obtained in this way
matched,  even roughly, the solid green line for the whole reconstructed field.
We therefore conclude that the gravitational coupling between angular modes plays a fundamental role in the onset of the observed instability.

We now proceed to examine whether the instability is dominantly tachyonic or powered by superradiance. The growth times, as shown in Fig.~\ref{fig:tacReconstructed}, can be as small as~$\sim 0.01 M$ . This seems to favor a tachyonic origin, as superradiance acts on longer timescales (see e.g.~\cite{Dima:2020rzg,Dolan:2007mj}). Moreover, the fact that the
instability is mostly due to the $m=0$ modes, and
that even the spherical mode $l=m=0$ can be unstable (see cyan long-dashed critical line in Fig.~\ref{fig:tacReconstructed}) bodes ill for superradiance, as these modes can never satisfy the superradiance condition  $\omega< m \Omega$ (with $\omega$ and $\Omega$ respectively the wave and horizon angular frequencies).

One may  naively expect the spherical mode $l=m=0$  not to suffer from a tachyonic instability either, since
 $\mu^2_\note{eff}=-\eta \mathcal{G}$ is  positive everywhere in a Schwarzschild spacetime when $\eta<0$ (as considered here).
However, the (squared) effective mass for the $l=m=0$ mode is  actually $-\eta \langle00| \mathcal{G}_{\rm Kerr}|00\rangle$,
which only matches the naive estimate $-\eta \mathcal{G}_{\rm Schwarzschild}$ at leading order in spin, correcting it by terms ${\cal O}(a^2)$.
This explains, in particular, why the spherical mode is stable at low spins.

To further confirm the tachyonic nature of the instabilities, we have conducted the following test. We re-ran our simulations with the (squared) effective mass replaced by its absolute value, $\mu^2_\note{eff} \rightarrow |\mu^2_\note{eff}|$. This is enough to suppress the instabilities, and further shows that the latter
were due to the change of sign of the GB invariant close to the horizon.
\begin{figure}[t]
    \includegraphics[scale=0.125,left]{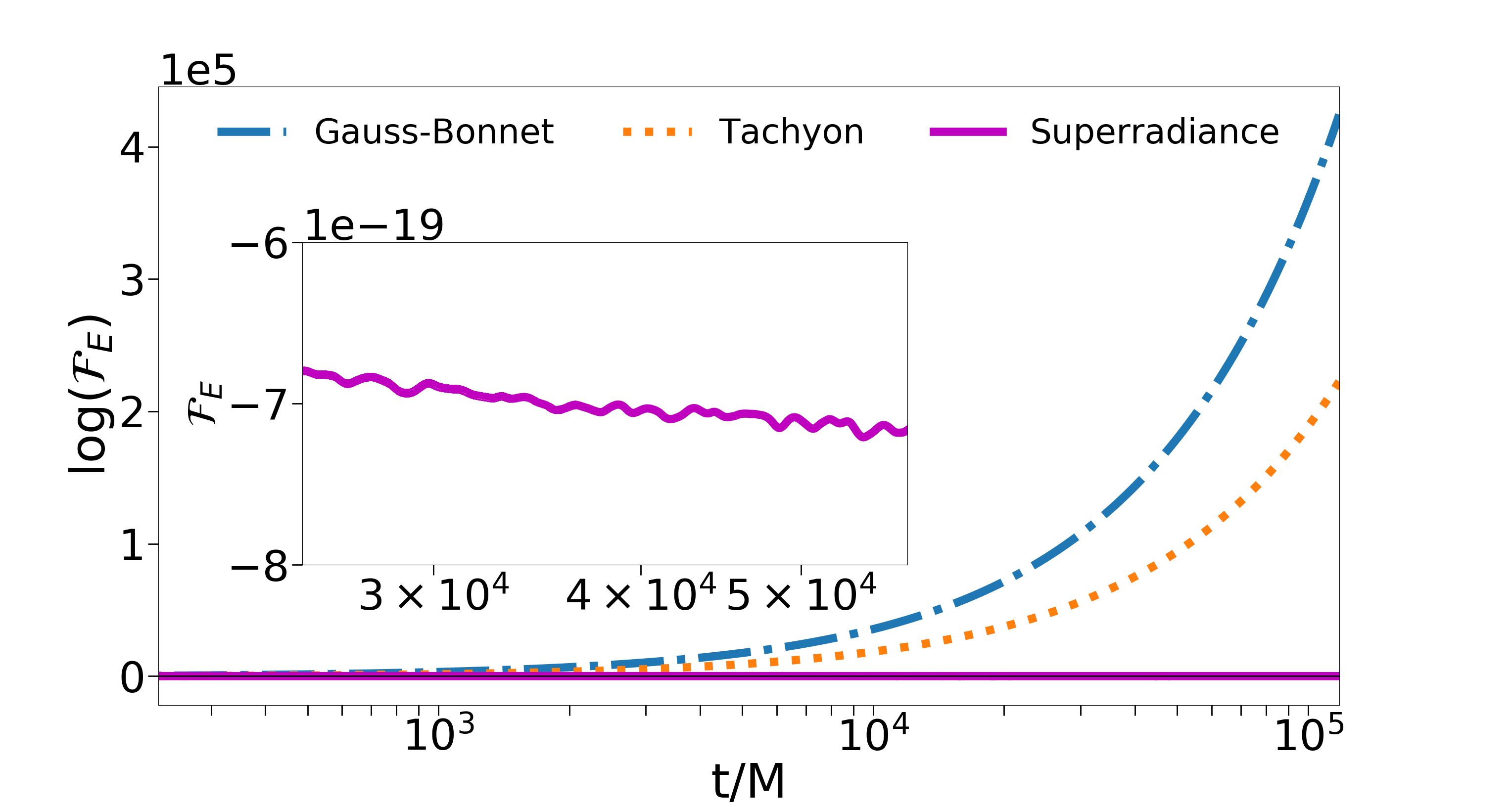}
    \caption{Energy flux $\mathcal{F}_E$ through the BH horizon vs  time, for $a=0.99 M$.
The blue, orange and magenta lines correspond respectively to $\eta=-10 M^2$, to a tachyonic mass $\mu M=i$, and to a constant, real mass
$\mu M=0.42$. The inset zooms on the constant, real mass flux (of which we show a moving average to decrease the oscillations caused by the dynamics). That flux is negative, signaling  energy extraction from the BH,
as expected for superradiant instabilities.}
    \label{fig:fluxes}
\end{figure}
One can also
look at the scalar fluxes through the event horizon
after the initial transient. In Fig.~\ref{fig:fluxes}, we compare the scalar field's energy flux through the horizon
for $\eta=-10 M^2$ (blue) vs the same fluxes for minimally coupled scalar fields with imaginary (orange) and real (magenta)
constant masses. Clearly, the flux for a scalar coupled to the GB invariant resembles  more closely the tachyonic (i.e. imaginary mass)
scalar field evolution, both in timescale and sign. Note  that the  constant, real mass case, whose
evolution is unstable due to superradiance, shows a slower  growth and negative energy fluxes. The latter
are indeed the hallmark of a superradiant instability, which {\it removes} rotational energy and angular
momentum from the BH.

The most plausible explanation for why Kerr BHs in qsGB  do not suffer from superradiant instabilities seems to be the
rapid falloff of the GB invariant (thus of the effective mass)
at large distances, $\mathcal{G}(r\rightarrow \infty) \sim 1/r^6$.
Scalar perturbations
with a position-dependent mass were  studied in~\cite{Dima:2020rzg}, which showed that
a  steep decay of the mass with distance quenches the superradiant instability. This happens
because the effective potential for scalar perturbations does not develop wells, and thus quasi-bound states,
unless the mass remains relatively constant till at least $r\sim 2-3 M$~\cite{Dima:2020rzg}.

\noindent\textbf{\textit{Conclusions:}} We have shown that a coupling, with a suitable sign, between a scalar and the GB invariant  can lead to an instability triggered by rapid rotation. We have also demonstrated that this instability is not related to superradiance, but is instead tachyonic in nature.
Nonlinear effects, which our approach does not capture, are expected to quench that instability and lead to a BH with scalar hair. The process is analogous to the more conventional spontaneous scalarization, but the threshold is controlled by the black hole rotation instead of its curvature.

The action that we use is sufficient for studying the onset of the instability that we have found for BHs. However, the endpoint of this instability, and hence the amount of hair a BH would carry, will strongly depend on nonlinear (self)interactions.\footnote{Stationary scalarized black hole solutions that constitute the endpoint of the instabilty will be presented elsewhere~\cite{Herdeiro:2020wei,Berti:2020kgk}.}  There is no obvious reason to believe that this instability is restricted to BHs, and it could well affect rapidly rotating stars as well. Hence, our results demonstrate that there is a broad class of theories where rotation might control deviations from GR. Our findings also have clear implications for searches of new physics in the strong-field regime.
Black hole scalar hair induces vacuum dipole gravitational emission, which is potentially observable
in the low frequency inspiral of binary system by gravitational wave interferometers~\cite{Barausse:2016eii,Toubiana:2020vtf},
deviations from GR in the spectrum of the gravitational wave ringdown~\cite{Berti:2016lat} or in the electromagnetic spectrum of
 accretion disks~\cite{Bambi:2011jq}, and
it may also impact the black hole shadow observed by the Event Horizon Telescope~\cite{Akiyama:2019cqa}.

We stress that we are not aware of any  observational upper bounds on $\eta$, which
we therefore allow here to reach very high values, for illustrative purposes and in order to excite higher modes.
Note that slowly rotating black holes in qsGB would be identical to their GR counterpart. Compact stars can scalarize for $\eta<0$
\cite{Silva:2017uqg} and hence yield constraints. However, this effect could easily be quenched by adding a coupling between the scalar field and the Ricci scalar \cite{Andreou:2019ikc,Ventagli:2020rnx}. The latter
might be necessary to get a sensible cosmology \cite{Antoniou:2020nax}, and
 would have no effect for black holes, thus leaving our analysis unaffected.

\begin{acknowledgments}

\noindent{{\bf{\em Acknowledgments.}}}
We thank Carlos Palenzuela and Miguel Bezares for insightful advice on technical aspects of the IMEX schemes and their validation,
and  Hector O. Silva for useful discussions on black hole  scalarization.
A.D., E.B. and N.F. acknowledge financial support provided under the European Union's H2020 ERC Consolidator Grant
``GRavity from Astrophysical to Microscopic Scales'' grant agreement no. GRAMS-815673.
T.P.S. acknowledges partial support from the STFC Consolidated Grant No.~ST/P000703/1.
We also acknowledge networking support by the COST Action GWverse Grant No.~CA16104.

\end{acknowledgments}

\bibliographystyle{apsrev4-1}
\bibliography{master}

\begin{thebibliography}{61}%
\makeatletter
\providecommand \@ifxundefined [1]{%
 \@ifx{#1\undefined}
}%
\providecommand \@ifnum [1]{%
 \ifnum #1\expandafter \@firstoftwo
 \else \expandafter \@secondoftwo
 \fi
}%
\providecommand \@ifx [1]{%
 \ifx #1\expandafter \@firstoftwo
 \else \expandafter \@secondoftwo
 \fi
}%
\providecommand \natexlab [1]{#1}%
\providecommand \enquote  [1]{``#1''}%
\providecommand \bibnamefont  [1]{#1}%
\providecommand \bibfnamefont [1]{#1}%
\providecommand \citenamefont [1]{#1}%
\providecommand \href@noop [0]{\@secondoftwo}%
\providecommand \href [0]{\begingroup \@sanitize@url \@href}%
\providecommand \@href[1]{\@@startlink{#1}\@@href}%
\providecommand \@@href[1]{\endgroup#1\@@endlink}%
\providecommand \@sanitize@url [0]{\catcode `\\12\catcode `\$12\catcode
  `\&12\catcode `\#12\catcode `\^12\catcode `\_12\catcode `\%12\relax}%
\providecommand \@@startlink[1]{}%
\providecommand \@@endlink[0]{}%
\providecommand \url  [0]{\begingroup\@sanitize@url \@url }%
\providecommand \@url [1]{\endgroup\@href {#1}{\urlprefix }}%
\providecommand \urlprefix  [0]{URL }%
\providecommand \Eprint [0]{\href }%
\providecommand \doibase [0]{http://dx.doi.org/}%
\providecommand \selectlanguage [0]{\@gobble}%
\providecommand \bibinfo  [0]{\@secondoftwo}%
\providecommand \bibfield  [0]{\@secondoftwo}%
\providecommand \translation [1]{[#1]}%
\providecommand \BibitemOpen [0]{}%
\providecommand \bibitemStop [0]{}%
\providecommand \bibitemNoStop [0]{.\EOS\space}%
\providecommand \EOS [0]{\spacefactor3000\relax}%
\providecommand \BibitemShut  [1]{\csname bibitem#1\endcsname}%
\let\auto@bib@innerbib\@empty
\bibitem [{\citenamefont {Webster}\ and\ \citenamefont
  {Murdin}(1972)}]{Webster:1972bsw}%
  \BibitemOpen
  \bibfield  {author} {\bibinfo {author} {\bibfnamefont {B.~L.}\ \bibnamefont
  {Webster}}\ and\ \bibinfo {author} {\bibfnamefont {P.}~\bibnamefont
  {Murdin}},\ }\href {\doibase 10.1038/235037a0} {\bibfield  {journal}
  {\bibinfo  {journal} {Nature}\ }\textbf {\bibinfo {volume} {235}},\ \bibinfo
  {pages} {37} (\bibinfo {year} {1972})}\BibitemShut {NoStop}%
\bibitem [{\citenamefont {{Reid}}\ \emph {et~al.}(1999)\citenamefont {{Reid}},
  \citenamefont {{Readhead}}, \citenamefont {{Vermeulen}},\ and\ \citenamefont
  {{Treuhaft}}}]{1999ApJ...524..816R}%
  \BibitemOpen
  \bibfield  {author} {\bibinfo {author} {\bibfnamefont {M.~J.}\ \bibnamefont
  {{Reid}}}, \bibinfo {author} {\bibfnamefont {A.~C.~S.}\ \bibnamefont
  {{Readhead}}}, \bibinfo {author} {\bibfnamefont {R.~C.}\ \bibnamefont
  {{Vermeulen}}}, \ and\ \bibinfo {author} {\bibfnamefont {R.~N.}\ \bibnamefont
  {{Treuhaft}}},\ }\href {\doibase 10.1086/307855} {\bibfield  {journal}
  {\bibinfo  {journal} {\apj}\ }\textbf {\bibinfo {volume} {524}},\ \bibinfo
  {pages} {816} (\bibinfo {year} {1999})},\ \Eprint
  {http://arxiv.org/abs/astro-ph/9905075} {arXiv:astro-ph/9905075 [astro-ph]}
  \BibitemShut {NoStop}%
\bibitem [{\citenamefont {Schodel}\ \emph {et~al.}(2002)\citenamefont {Schodel}
  \emph {et~al.}}]{Schodel:2002vg}%
  \BibitemOpen
  \bibfield  {author} {\bibinfo {author} {\bibfnamefont {R.}~\bibnamefont
  {Schodel}} \emph {et~al.},\ }\href {\doibase 10.1038/nature01121} {\bibfield
  {journal} {\bibinfo  {journal} {Nature}\ }\textbf {\bibinfo {volume} {419}},\
  \bibinfo {pages} {694} (\bibinfo {year} {2002})},\ \Eprint
  {http://arxiv.org/abs/astro-ph/0210426} {arXiv:astro-ph/0210426 [astro-ph]}
  \BibitemShut {NoStop}%
\bibitem [{\citenamefont {Reid}\ \emph {et~al.}(2003)\citenamefont {Reid},
  \citenamefont {Menten}, \citenamefont {Genzel}, \citenamefont {Ott},
  \citenamefont {Schodel},\ and\ \citenamefont {Eckart}}]{Reid_2003}%
  \BibitemOpen
  \bibfield  {author} {\bibinfo {author} {\bibfnamefont {M.~J.}\ \bibnamefont
  {Reid}}, \bibinfo {author} {\bibfnamefont {K.~M.}\ \bibnamefont {Menten}},
  \bibinfo {author} {\bibfnamefont {R.}~\bibnamefont {Genzel}}, \bibinfo
  {author} {\bibfnamefont {T.}~\bibnamefont {Ott}}, \bibinfo {author}
  {\bibfnamefont {R.}~\bibnamefont {Schodel}}, \ and\ \bibinfo {author}
  {\bibfnamefont {A.}~\bibnamefont {Eckart}},\ }\href {\doibase 10.1086/368074}
  {\bibfield  {journal} {\bibinfo  {journal} {The Astrophysical Journal}\
  }\textbf {\bibinfo {volume} {587}},\ \bibinfo {pages} {208} (\bibinfo {year}
  {2003})}\BibitemShut {NoStop}%
\bibitem [{\citenamefont {Gillessen}\ \emph {et~al.}(2009)\citenamefont
  {Gillessen}, \citenamefont {Eisenhauer}, \citenamefont {Trippe},
  \citenamefont {Alexander}, \citenamefont {Genzel}, \citenamefont {Martins},\
  and\ \citenamefont {Ott}}]{Gillessen:2008qv}%
  \BibitemOpen
  \bibfield  {author} {\bibinfo {author} {\bibfnamefont {S.}~\bibnamefont
  {Gillessen}}, \bibinfo {author} {\bibfnamefont {F.}~\bibnamefont
  {Eisenhauer}}, \bibinfo {author} {\bibfnamefont {S.}~\bibnamefont {Trippe}},
  \bibinfo {author} {\bibfnamefont {T.}~\bibnamefont {Alexander}}, \bibinfo
  {author} {\bibfnamefont {R.}~\bibnamefont {Genzel}}, \bibinfo {author}
  {\bibfnamefont {F.}~\bibnamefont {Martins}}, \ and\ \bibinfo {author}
  {\bibfnamefont {T.}~\bibnamefont {Ott}},\ }\href {\doibase
  10.1088/0004-637X/692/2/1075} {\bibfield  {journal} {\bibinfo  {journal}
  {Astrophys. J.}\ }\textbf {\bibinfo {volume} {692}},\ \bibinfo {pages} {1075}
  (\bibinfo {year} {2009})},\ \Eprint {http://arxiv.org/abs/0810.4674}
  {arXiv:0810.4674 [astro-ph]} \BibitemShut {NoStop}%
\bibitem [{\citenamefont {Abbott}\ \emph {et~al.}(2016)\citenamefont {Abbott}
  \emph {et~al.}}]{Abbott:2016blz}%
  \BibitemOpen
  \bibfield  {author} {\bibinfo {author} {\bibfnamefont {B.~P.}\ \bibnamefont
  {Abbott}} \emph {et~al.} (\bibinfo {collaboration} {LIGO Scientific,
  Virgo}),\ }\href {\doibase 10.1103/PhysRevLett.116.061102} {\bibfield
  {journal} {\bibinfo  {journal} {Phys. Rev. Lett.}\ }\textbf {\bibinfo
  {volume} {116}},\ \bibinfo {pages} {061102} (\bibinfo {year} {2016})},\
  \Eprint {http://arxiv.org/abs/1602.03837} {arXiv:1602.03837 [gr-qc]}
  \BibitemShut {NoStop}%
\bibitem [{\citenamefont {Abuter}\ \emph {et~al.}(2018)\citenamefont {Abuter}
  \emph {et~al.}}]{2018A&A...618L..10G}%
  \BibitemOpen
  \bibfield  {author} {\bibinfo {author} {\bibfnamefont {R.}~\bibnamefont
  {Abuter}} \emph {et~al.} (\bibinfo {collaboration} {{Gravity
  Collaboration}}),\ }\href {\doibase 10.1051/0004-6361/201834294} {\bibfield
  {journal} {\bibinfo  {journal} {Astron. \& Astrophys.}\ }\textbf {\bibinfo
  {volume} {618}},\ \bibinfo {pages} {L10} (\bibinfo {year} {2018})},\ \Eprint
  {http://arxiv.org/abs/1810.12641} {arXiv:1810.12641 [astro-ph.GA]}
  \BibitemShut {NoStop}%
\bibitem [{\citenamefont {Akiyama}\ \emph {et~al.}(2019)\citenamefont {Akiyama}
  \emph {et~al.}}]{Akiyama:2019cqa}%
  \BibitemOpen
  \bibfield  {author} {\bibinfo {author} {\bibfnamefont {K.}~\bibnamefont
  {Akiyama}} \emph {et~al.} (\bibinfo {collaboration} {Event Horizon
  Telescope}),\ }\href {\doibase 10.3847/2041-8213/ab0ec7} {\bibfield
  {journal} {\bibinfo  {journal} {Astrophys. J.}\ }\textbf {\bibinfo {volume}
  {875}},\ \bibinfo {pages} {L1} (\bibinfo {year} {2019})},\ \Eprint
  {http://arxiv.org/abs/1906.11238} {arXiv:1906.11238 [astro-ph.GA]}
  \BibitemShut {NoStop}%
\bibitem [{\citenamefont {Barausse}\ \emph {et~al.}(2014)\citenamefont
  {Barausse}, \citenamefont {Cardoso},\ and\ \citenamefont
  {Pani}}]{Barausse:2014tra}%
  \BibitemOpen
  \bibfield  {author} {\bibinfo {author} {\bibfnamefont {E.}~\bibnamefont
  {Barausse}}, \bibinfo {author} {\bibfnamefont {V.}~\bibnamefont {Cardoso}}, \
  and\ \bibinfo {author} {\bibfnamefont {P.}~\bibnamefont {Pani}},\ }\href
  {\doibase 10.1103/PhysRevD.89.104059} {\bibfield  {journal} {\bibinfo
  {journal} {Phys. Rev.}\ }\textbf {\bibinfo {volume} {D89}},\ \bibinfo {pages}
  {104059} (\bibinfo {year} {2014})},\ \Eprint {http://arxiv.org/abs/1404.7149}
  {arXiv:1404.7149 [gr-qc]} \BibitemShut {NoStop}%
\bibitem [{\citenamefont {Israel}(1967)}]{Israel:1967wq}%
  \BibitemOpen
  \bibfield  {author} {\bibinfo {author} {\bibfnamefont {W.}~\bibnamefont
  {Israel}},\ }\href {\doibase 10.1103/PhysRev.164.1776} {\bibfield  {journal}
  {\bibinfo  {journal} {Phys. Rev.}\ }\textbf {\bibinfo {volume} {164}},\
  \bibinfo {pages} {1776} (\bibinfo {year} {1967})}\BibitemShut {NoStop}%
\bibitem [{\citenamefont {Carter}(1971)}]{Carter:1971zc}%
  \BibitemOpen
  \bibfield  {author} {\bibinfo {author} {\bibfnamefont {B.}~\bibnamefont
  {Carter}},\ }\href {\doibase 10.1103/PhysRevLett.26.331} {\bibfield
  {journal} {\bibinfo  {journal} {Phys. Rev. Lett.}\ }\textbf {\bibinfo
  {volume} {26}},\ \bibinfo {pages} {331} (\bibinfo {year} {1971})}\BibitemShut
  {NoStop}%
\bibitem [{\citenamefont {Robinson}(1975)}]{Robinson:1975bv}%
  \BibitemOpen
  \bibfield  {author} {\bibinfo {author} {\bibfnamefont {D.~C.}\ \bibnamefont
  {Robinson}},\ }\href {\doibase 10.1103/PhysRevLett.34.905} {\bibfield
  {journal} {\bibinfo  {journal} {Phys. Rev. Lett.}\ }\textbf {\bibinfo
  {volume} {34}},\ \bibinfo {pages} {905} (\bibinfo {year} {1975})}\BibitemShut
  {NoStop}%
\bibitem [{\citenamefont {Berti}\ \emph {et~al.}(2016)\citenamefont {Berti},
  \citenamefont {Sesana}, \citenamefont {Barausse}, \citenamefont {Cardoso},\
  and\ \citenamefont {Belczynski}}]{Berti:2016lat}%
  \BibitemOpen
  \bibfield  {author} {\bibinfo {author} {\bibfnamefont {E.}~\bibnamefont
  {Berti}}, \bibinfo {author} {\bibfnamefont {A.}~\bibnamefont {Sesana}},
  \bibinfo {author} {\bibfnamefont {E.}~\bibnamefont {Barausse}}, \bibinfo
  {author} {\bibfnamefont {V.}~\bibnamefont {Cardoso}}, \ and\ \bibinfo
  {author} {\bibfnamefont {K.}~\bibnamefont {Belczynski}},\ }\href {\doibase
  10.1103/PhysRevLett.117.101102} {\bibfield  {journal} {\bibinfo  {journal}
  {Phys.\ Rev.\ Lett.}\ }\textbf {\bibinfo {volume} {117}},\ \bibinfo {pages}
  {101102} (\bibinfo {year} {2016})},\ \Eprint
  {http://arxiv.org/abs/1605.09286} {arXiv:1605.09286 [gr-qc]} \BibitemShut
  {NoStop}%
\bibitem [{\citenamefont {Barausse}\ \emph {et~al.}(2016)\citenamefont
  {Barausse}, \citenamefont {Yunes},\ and\ \citenamefont
  {Chamberlain}}]{Barausse:2016eii}%
  \BibitemOpen
  \bibfield  {author} {\bibinfo {author} {\bibfnamefont {E.}~\bibnamefont
  {Barausse}}, \bibinfo {author} {\bibfnamefont {N.}~\bibnamefont {Yunes}}, \
  and\ \bibinfo {author} {\bibfnamefont {K.}~\bibnamefont {Chamberlain}},\
  }\href {\doibase 10.1103/PhysRevLett.116.241104} {\bibfield  {journal}
  {\bibinfo  {journal} {Phys.\ Rev.\ Lett.}\ }\textbf {\bibinfo {volume}
  {116}},\ \bibinfo {pages} {241104} (\bibinfo {year} {2016})},\ \Eprint
  {http://arxiv.org/abs/1603.04075} {arXiv:1603.04075 [gr-qc]} \BibitemShut
  {NoStop}%
\bibitem [{\citenamefont {Toubiana}\ \emph {et~al.}(2020)\citenamefont
  {Toubiana}, \citenamefont {Marsat}, \citenamefont {Babak}, \citenamefont
  {Barausse},\ and\ \citenamefont {Baker}}]{Toubiana:2020vtf}%
  \BibitemOpen
  \bibfield  {author} {\bibinfo {author} {\bibfnamefont {A.}~\bibnamefont
  {Toubiana}}, \bibinfo {author} {\bibfnamefont {S.}~\bibnamefont {Marsat}},
  \bibinfo {author} {\bibfnamefont {S.}~\bibnamefont {Babak}}, \bibinfo
  {author} {\bibfnamefont {E.}~\bibnamefont {Barausse}}, \ and\ \bibinfo
  {author} {\bibfnamefont {J.}~\bibnamefont {Baker}},\ }\href {\doibase
  10.1103/PhysRevD.101.104038} {\bibfield  {journal} {\bibinfo  {journal}
  {Phys. Rev. D}\ }\textbf {\bibinfo {volume} {101}},\ \bibinfo {pages}
  {104038} (\bibinfo {year} {2020})},\ \Eprint
  {http://arxiv.org/abs/2004.03626} {arXiv:2004.03626 [gr-qc]} \BibitemShut
  {NoStop}%
\bibitem [{\citenamefont {Abbott}\ \emph {et~al.}(2019)\citenamefont {Abbott}
  \emph {et~al.}}]{LIGOScientific:2019fpa}%
  \BibitemOpen
  \bibfield  {author} {\bibinfo {author} {\bibfnamefont {B.~P.}\ \bibnamefont
  {Abbott}} \emph {et~al.} (\bibinfo {collaboration} {LIGO Scientific,
  Virgo}),\ }\href {\doibase 10.1103/PhysRevD.100.104036} {\bibfield  {journal}
  {\bibinfo  {journal} {Phys. Rev.}\ }\textbf {\bibinfo {volume} {D100}},\
  \bibinfo {pages} {104036} (\bibinfo {year} {2019})},\ \Eprint
  {http://arxiv.org/abs/1903.04467} {arXiv:1903.04467 [gr-qc]} \BibitemShut
  {NoStop}%
\bibitem [{\citenamefont {Isi}\ \emph {et~al.}(2019)\citenamefont {Isi},
  \citenamefont {Giesler}, \citenamefont {Farr}, \citenamefont {Scheel},\ and\
  \citenamefont {Teukolsky}}]{Isi:2019aib}%
  \BibitemOpen
  \bibfield  {author} {\bibinfo {author} {\bibfnamefont {M.}~\bibnamefont
  {Isi}}, \bibinfo {author} {\bibfnamefont {M.}~\bibnamefont {Giesler}},
  \bibinfo {author} {\bibfnamefont {W.~M.}\ \bibnamefont {Farr}}, \bibinfo
  {author} {\bibfnamefont {M.~A.}\ \bibnamefont {Scheel}}, \ and\ \bibinfo
  {author} {\bibfnamefont {S.~A.}\ \bibnamefont {Teukolsky}},\ }\href {\doibase
  10.1103/PhysRevLett.123.111102} {\bibfield  {journal} {\bibinfo  {journal}
  {Phys. Rev. Lett.}\ }\textbf {\bibinfo {volume} {123}},\ \bibinfo {pages}
  {111102} (\bibinfo {year} {2019})},\ \Eprint
  {http://arxiv.org/abs/1905.00869} {arXiv:1905.00869 [gr-qc]} \BibitemShut
  {NoStop}%
\bibitem [{\citenamefont {Hawking}(1972)}]{Hawking:1972qk}%
  \BibitemOpen
  \bibfield  {author} {\bibinfo {author} {\bibfnamefont {S.~W.}\ \bibnamefont
  {Hawking}},\ }\href {\doibase 10.1007/BF01877518} {\bibfield  {journal}
  {\bibinfo  {journal} {Commun. Math. Phys.}\ }\textbf {\bibinfo {volume}
  {25}},\ \bibinfo {pages} {167} (\bibinfo {year} {1972})}\BibitemShut
  {NoStop}%
\bibitem [{\citenamefont {Sotiriou}\ and\ \citenamefont
  {Faraoni}(2012)}]{Sotiriou:2011dz}%
  \BibitemOpen
  \bibfield  {author} {\bibinfo {author} {\bibfnamefont {T.~P.}\ \bibnamefont
  {Sotiriou}}\ and\ \bibinfo {author} {\bibfnamefont {V.}~\bibnamefont
  {Faraoni}},\ }\href {\doibase 10.1103/PhysRevLett.108.081103} {\bibfield
  {journal} {\bibinfo  {journal} {Phys. Rev. Lett.}\ }\textbf {\bibinfo
  {volume} {108}},\ \bibinfo {pages} {081103} (\bibinfo {year} {2012})},\
  \Eprint {http://arxiv.org/abs/1109.6324} {arXiv:1109.6324 [gr-qc]}
  \BibitemShut {NoStop}%
\bibitem [{\citenamefont {Hui}\ and\ \citenamefont
  {Nicolis}(2013)}]{Hui:2012qt}%
  \BibitemOpen
  \bibfield  {author} {\bibinfo {author} {\bibfnamefont {L.}~\bibnamefont
  {Hui}}\ and\ \bibinfo {author} {\bibfnamefont {A.}~\bibnamefont {Nicolis}},\
  }\href {\doibase 10.1103/PhysRevLett.110.241104} {\bibfield  {journal}
  {\bibinfo  {journal} {Phys. Rev. Lett.}\ }\textbf {\bibinfo {volume} {110}},\
  \bibinfo {pages} {241104} (\bibinfo {year} {2013})},\ \Eprint
  {http://arxiv.org/abs/1202.1296} {arXiv:1202.1296 [hep-th]} \BibitemShut
  {NoStop}%
\bibitem [{\citenamefont {Sotiriou}\ and\ \citenamefont
  {Zhou}(2014{\natexlab{a}})}]{Sotiriou:2013qea}%
  \BibitemOpen
  \bibfield  {author} {\bibinfo {author} {\bibfnamefont {T.~P.}\ \bibnamefont
  {Sotiriou}}\ and\ \bibinfo {author} {\bibfnamefont {S.-Y.}\ \bibnamefont
  {Zhou}},\ }\href {\doibase 10.1103/PhysRevLett.112.251102} {\bibfield
  {journal} {\bibinfo  {journal} {Phys. Rev. Lett.}\ }\textbf {\bibinfo
  {volume} {112}},\ \bibinfo {pages} {251102} (\bibinfo {year}
  {2014}{\natexlab{a}})},\ \Eprint {http://arxiv.org/abs/1312.3622}
  {arXiv:1312.3622 [gr-qc]} \BibitemShut {NoStop}%
\bibitem [{\citenamefont {Barausse}\ and\ \citenamefont
  {Yagi}(2015)}]{starsNohair1}%
  \BibitemOpen
  \bibfield  {author} {\bibinfo {author} {\bibfnamefont {E.}~\bibnamefont
  {Barausse}}\ and\ \bibinfo {author} {\bibfnamefont {K.}~\bibnamefont
  {Yagi}},\ }\href {\doibase 10.1103/PhysRevLett.115.211105} {\bibfield
  {journal} {\bibinfo  {journal} {Phys. Rev. Lett.}\ }\textbf {\bibinfo
  {volume} {115}},\ \bibinfo {pages} {211105} (\bibinfo {year} {2015})},\
  \Eprint {http://arxiv.org/abs/1509.04539} {arXiv:1509.04539 [gr-qc]}
  \BibitemShut {NoStop}%
\bibitem [{\citenamefont {Barausse}(2017)}]{starsNohair2}%
  \BibitemOpen
  \bibfield  {author} {\bibinfo {author} {\bibfnamefont {E.}~\bibnamefont
  {Barausse}},\ }\bibfield  {booktitle} {\emph {\bibinfo {booktitle}
  {{Proceedings, 3rd International Symposium on Quest for the Origin of
  Particles and the Universe (KMI2017): Nagoya, Japan, January 5-7, 2017}}},\
  }\href {\doibase 10.22323/1.294.0029} {\bibfield  {journal} {\bibinfo
  {journal} {PoS}\ }\textbf {\bibinfo {volume} {KMI2017}},\ \bibinfo {pages}
  {029} (\bibinfo {year} {2017})},\ \Eprint {http://arxiv.org/abs/1703.05699}
  {arXiv:1703.05699 [gr-qc]} \BibitemShut {NoStop}%
\bibitem [{\citenamefont {Lehebel}\ \emph {et~al.}(2017)\citenamefont
  {Lehebel}, \citenamefont {Babichev},\ and\ \citenamefont
  {Charmousis}}]{Lehebel:2017fag}%
  \BibitemOpen
  \bibfield  {author} {\bibinfo {author} {\bibfnamefont {A.}~\bibnamefont
  {Lehebel}}, \bibinfo {author} {\bibfnamefont {E.}~\bibnamefont {Babichev}}, \
  and\ \bibinfo {author} {\bibfnamefont {C.}~\bibnamefont {Charmousis}},\
  }\href {\doibase 10.1088/1475-7516/2017/07/037} {\bibfield  {journal}
  {\bibinfo  {journal} {JCAP}\ }\textbf {\bibinfo {volume} {07}},\ \bibinfo
  {pages} {037} (\bibinfo {year} {2017})},\ \Eprint
  {http://arxiv.org/abs/1706.04989} {arXiv:1706.04989 [gr-qc]} \BibitemShut
  {NoStop}%
\bibitem [{\citenamefont {Yagi}\ \emph {et~al.}(2016)\citenamefont {Yagi},
  \citenamefont {Stein},\ and\ \citenamefont {Yunes}}]{Yagi:2015oca}%
  \BibitemOpen
  \bibfield  {author} {\bibinfo {author} {\bibfnamefont {K.}~\bibnamefont
  {Yagi}}, \bibinfo {author} {\bibfnamefont {L.~C.}\ \bibnamefont {Stein}}, \
  and\ \bibinfo {author} {\bibfnamefont {N.}~\bibnamefont {Yunes}},\ }\href
  {\doibase 10.1103/PhysRevD.93.024010} {\bibfield  {journal} {\bibinfo
  {journal} {Phys. Rev. D}\ }\textbf {\bibinfo {volume} {93}},\ \bibinfo
  {pages} {024010} (\bibinfo {year} {2016})},\ \Eprint
  {http://arxiv.org/abs/1510.02152} {arXiv:1510.02152 [gr-qc]} \BibitemShut
  {NoStop}%
\bibitem [{\citenamefont {Sotiriou}\ and\ \citenamefont
  {Zhou}(2014{\natexlab{b}})}]{Sotiriou:2014pfa}%
  \BibitemOpen
  \bibfield  {author} {\bibinfo {author} {\bibfnamefont {T.~P.}\ \bibnamefont
  {Sotiriou}}\ and\ \bibinfo {author} {\bibfnamefont {S.-Y.}\ \bibnamefont
  {Zhou}},\ }\href {\doibase 10.1103/PhysRevD.90.124063} {\bibfield  {journal}
  {\bibinfo  {journal} {Phys. Rev.}\ }\textbf {\bibinfo {volume} {D90}},\
  \bibinfo {pages} {124063} (\bibinfo {year} {2014}{\natexlab{b}})},\ \Eprint
  {http://arxiv.org/abs/1408.1698} {arXiv:1408.1698 [gr-qc]} \BibitemShut
  {NoStop}%
\bibitem [{\citenamefont {Babichev}\ and\ \citenamefont
  {Charmousis}(2014)}]{Babichev:2013cya}%
  \BibitemOpen
  \bibfield  {author} {\bibinfo {author} {\bibfnamefont {E.}~\bibnamefont
  {Babichev}}\ and\ \bibinfo {author} {\bibfnamefont {C.}~\bibnamefont
  {Charmousis}},\ }\href {\doibase 10.1007/JHEP08(2014)106} {\bibfield
  {journal} {\bibinfo  {journal} {JHEP}\ }\textbf {\bibinfo {volume} {08}},\
  \bibinfo {pages} {106} (\bibinfo {year} {2014})},\ \Eprint
  {http://arxiv.org/abs/1312.3204} {arXiv:1312.3204 [gr-qc]} \BibitemShut
  {NoStop}%
\bibitem [{\citenamefont {Cardoso}\ \emph
  {et~al.}(2013{\natexlab{a}})\citenamefont {Cardoso}, \citenamefont {Carucci},
  \citenamefont {Pani},\ and\ \citenamefont {Sotiriou}}]{Cardoso:2013fwa}%
  \BibitemOpen
  \bibfield  {author} {\bibinfo {author} {\bibfnamefont {V.}~\bibnamefont
  {Cardoso}}, \bibinfo {author} {\bibfnamefont {I.~P.}\ \bibnamefont
  {Carucci}}, \bibinfo {author} {\bibfnamefont {P.}~\bibnamefont {Pani}}, \
  and\ \bibinfo {author} {\bibfnamefont {T.~P.}\ \bibnamefont {Sotiriou}},\
  }\href {\doibase 10.1103/PhysRevLett.111.111101} {\bibfield  {journal}
  {\bibinfo  {journal} {Phys. Rev. Lett.}\ }\textbf {\bibinfo {volume} {111}},\
  \bibinfo {pages} {111101} (\bibinfo {year} {2013}{\natexlab{a}})},\ \Eprint
  {http://arxiv.org/abs/1308.6587} {arXiv:1308.6587 [gr-qc]} \BibitemShut
  {NoStop}%
\bibitem [{\citenamefont {Cardoso}\ \emph
  {et~al.}(2013{\natexlab{b}})\citenamefont {Cardoso}, \citenamefont {Carucci},
  \citenamefont {Pani},\ and\ \citenamefont {Sotiriou}}]{Cardoso:2013opa}%
  \BibitemOpen
  \bibfield  {author} {\bibinfo {author} {\bibfnamefont {V.}~\bibnamefont
  {Cardoso}}, \bibinfo {author} {\bibfnamefont {I.~P.}\ \bibnamefont
  {Carucci}}, \bibinfo {author} {\bibfnamefont {P.}~\bibnamefont {Pani}}, \
  and\ \bibinfo {author} {\bibfnamefont {T.~P.}\ \bibnamefont {Sotiriou}},\
  }\href {\doibase 10.1103/PhysRevD.88.044056} {\bibfield  {journal} {\bibinfo
  {journal} {Phys. Rev.}\ }\textbf {\bibinfo {volume} {D88}},\ \bibinfo {pages}
  {044056} (\bibinfo {year} {2013}{\natexlab{b}})},\ \Eprint
  {http://arxiv.org/abs/1305.6936} {arXiv:1305.6936 [gr-qc]} \BibitemShut
  {NoStop}%
\bibitem [{\citenamefont {Herdeiro}\ and\ \citenamefont
  {Radu}(2014)}]{Herdeiro:2014goa}%
  \BibitemOpen
  \bibfield  {author} {\bibinfo {author} {\bibfnamefont {C.~A.~R.}\
  \bibnamefont {Herdeiro}}\ and\ \bibinfo {author} {\bibfnamefont
  {E.}~\bibnamefont {Radu}},\ }\href {\doibase 10.1103/PhysRevLett.112.221101}
  {\bibfield  {journal} {\bibinfo  {journal} {Phys. Rev. Lett.}\ }\textbf
  {\bibinfo {volume} {112}},\ \bibinfo {pages} {221101} (\bibinfo {year}
  {2014})},\ \Eprint {http://arxiv.org/abs/1403.2757} {arXiv:1403.2757 [gr-qc]}
  \BibitemShut {NoStop}%
\bibitem [{\citenamefont {Antoniou}\ \emph {et~al.}(2018)\citenamefont
  {Antoniou}, \citenamefont {Bakopoulos},\ and\ \citenamefont
  {Kanti}}]{Antoniou:2017acq}%
  \BibitemOpen
  \bibfield  {author} {\bibinfo {author} {\bibfnamefont {G.}~\bibnamefont
  {Antoniou}}, \bibinfo {author} {\bibfnamefont {A.}~\bibnamefont
  {Bakopoulos}}, \ and\ \bibinfo {author} {\bibfnamefont {P.}~\bibnamefont
  {Kanti}},\ }\href {\doibase 10.1103/PhysRevLett.120.131102} {\bibfield
  {journal} {\bibinfo  {journal} {Phys. Rev. Lett.}\ }\textbf {\bibinfo
  {volume} {120}},\ \bibinfo {pages} {131102} (\bibinfo {year} {2018})},\
  \Eprint {http://arxiv.org/abs/1711.03390} {arXiv:1711.03390 [hep-th]}
  \BibitemShut {NoStop}%
\bibitem [{\citenamefont {Sotiriou}(2015)}]{Sotiriou:2015pka}%
  \BibitemOpen
  \bibfield  {author} {\bibinfo {author} {\bibfnamefont {T.~P.}\ \bibnamefont
  {Sotiriou}},\ }\href {\doibase 10.1088/0264-9381/32/21/214002} {\bibfield
  {journal} {\bibinfo  {journal} {Class. Quant. Grav.}\ }\textbf {\bibinfo
  {volume} {32}},\ \bibinfo {pages} {214002} (\bibinfo {year} {2015})},\
  \Eprint {http://arxiv.org/abs/1505.00248} {arXiv:1505.00248 [gr-qc]}
  \BibitemShut {NoStop}%
\bibitem [{\citenamefont {Silva}\ \emph {et~al.}(2018)\citenamefont {Silva},
  \citenamefont {Sakstein}, \citenamefont {Gualtieri}, \citenamefont
  {Sotiriou},\ and\ \citenamefont {Berti}}]{Silva:2017uqg}%
  \BibitemOpen
  \bibfield  {author} {\bibinfo {author} {\bibfnamefont {H.~O.}\ \bibnamefont
  {Silva}}, \bibinfo {author} {\bibfnamefont {J.}~\bibnamefont {Sakstein}},
  \bibinfo {author} {\bibfnamefont {L.}~\bibnamefont {Gualtieri}}, \bibinfo
  {author} {\bibfnamefont {T.~P.}\ \bibnamefont {Sotiriou}}, \ and\ \bibinfo
  {author} {\bibfnamefont {E.}~\bibnamefont {Berti}},\ }\href {\doibase
  10.1103/PhysRevLett.120.131104} {\bibfield  {journal} {\bibinfo  {journal}
  {Phys. Rev. Lett.}\ }\textbf {\bibinfo {volume} {120}},\ \bibinfo {pages}
  {131104} (\bibinfo {year} {2018})},\ \Eprint
  {http://arxiv.org/abs/1711.02080} {arXiv:1711.02080 [gr-qc]} \BibitemShut
  {NoStop}%
\bibitem [{\citenamefont {Doneva}\ and\ \citenamefont
  {Yazadjiev}(2018)}]{Doneva:2017bvd}%
  \BibitemOpen
  \bibfield  {author} {\bibinfo {author} {\bibfnamefont {D.~D.}\ \bibnamefont
  {Doneva}}\ and\ \bibinfo {author} {\bibfnamefont {S.~S.}\ \bibnamefont
  {Yazadjiev}},\ }\href {\doibase 10.1103/PhysRevLett.120.131103} {\bibfield
  {journal} {\bibinfo  {journal} {Phys. Rev. Lett.}\ }\textbf {\bibinfo
  {volume} {120}},\ \bibinfo {pages} {131103} (\bibinfo {year} {2018})},\
  \Eprint {http://arxiv.org/abs/1711.01187} {arXiv:1711.01187 [gr-qc]}
  \BibitemShut {NoStop}%
\bibitem [{\citenamefont {Barausse}\ and\ \citenamefont
  {Sotiriou}(2008)}]{Barausse:2008xv}%
  \BibitemOpen
  \bibfield  {author} {\bibinfo {author} {\bibfnamefont {E.}~\bibnamefont
  {Barausse}}\ and\ \bibinfo {author} {\bibfnamefont {T.~P.}\ \bibnamefont
  {Sotiriou}},\ }\href {\doibase 10.1103/PhysRevLett.101.099001} {\bibfield
  {journal} {\bibinfo  {journal} {Phys. Rev. Lett.}\ }\textbf {\bibinfo
  {volume} {101}},\ \bibinfo {pages} {099001} (\bibinfo {year} {2008})},\
  \Eprint {http://arxiv.org/abs/0803.3433} {arXiv:0803.3433 [gr-qc]}
  \BibitemShut {NoStop}%
\bibitem [{\citenamefont {Damour}\ and\ \citenamefont
  {Esposito-Farese}(1993)}]{Damour:1993hw}%
  \BibitemOpen
  \bibfield  {author} {\bibinfo {author} {\bibfnamefont {T.}~\bibnamefont
  {Damour}}\ and\ \bibinfo {author} {\bibfnamefont {G.}~\bibnamefont
  {Esposito-Farese}},\ }\href {\doibase 10.1103/PhysRevLett.70.2220} {\bibfield
   {journal} {\bibinfo  {journal} {Phys. Rev. Lett.}\ }\textbf {\bibinfo
  {volume} {70}},\ \bibinfo {pages} {2220} (\bibinfo {year}
  {1993})}\BibitemShut {NoStop}%
\bibitem [{\citenamefont {Barausse}\ \emph {et~al.}(2013)\citenamefont
  {Barausse}, \citenamefont {Palenzuela}, \citenamefont {Ponce},\ and\
  \citenamefont {Lehner}}]{ST1}%
  \BibitemOpen
  \bibfield  {author} {\bibinfo {author} {\bibfnamefont {E.}~\bibnamefont
  {Barausse}}, \bibinfo {author} {\bibfnamefont {C.}~\bibnamefont
  {Palenzuela}}, \bibinfo {author} {\bibfnamefont {M.}~\bibnamefont {Ponce}}, \
  and\ \bibinfo {author} {\bibfnamefont {L.}~\bibnamefont {Lehner}},\ }\href
  {\doibase 10.1103/PhysRevD.87.081506} {\bibfield  {journal} {\bibinfo
  {journal} {Phys. Rev. D}\ }\textbf {\bibinfo {volume} {87}},\ \bibinfo
  {pages} {081506} (\bibinfo {year} {2013})}\BibitemShut {NoStop}%
\bibitem [{\citenamefont {Palenzuela}\ \emph {et~al.}(2014)\citenamefont
  {Palenzuela}, \citenamefont {Barausse}, \citenamefont {Ponce},\ and\
  \citenamefont {Lehner}}]{ST2}%
  \BibitemOpen
  \bibfield  {author} {\bibinfo {author} {\bibfnamefont {C.}~\bibnamefont
  {Palenzuela}}, \bibinfo {author} {\bibfnamefont {E.}~\bibnamefont
  {Barausse}}, \bibinfo {author} {\bibfnamefont {M.}~\bibnamefont {Ponce}}, \
  and\ \bibinfo {author} {\bibfnamefont {L.}~\bibnamefont {Lehner}},\ }\href
  {\doibase 10.1103/PhysRevD.89.044024} {\bibfield  {journal} {\bibinfo
  {journal} {Phys. Rev. D}\ }\textbf {\bibinfo {volume} {89}},\ \bibinfo
  {pages} {044024} (\bibinfo {year} {2014})}\BibitemShut {NoStop}%
\bibitem [{\citenamefont {Shibata}\ \emph {et~al.}(2014)\citenamefont
  {Shibata}, \citenamefont {Taniguchi}, \citenamefont {Okawa},\ and\
  \citenamefont {Buonanno}}]{ST3}%
  \BibitemOpen
  \bibfield  {author} {\bibinfo {author} {\bibfnamefont {M.}~\bibnamefont
  {Shibata}}, \bibinfo {author} {\bibfnamefont {K.}~\bibnamefont {Taniguchi}},
  \bibinfo {author} {\bibfnamefont {H.}~\bibnamefont {Okawa}}, \ and\ \bibinfo
  {author} {\bibfnamefont {A.}~\bibnamefont {Buonanno}},\ }\href {\doibase
  10.1103/PhysRevD.89.084005} {\bibfield  {journal} {\bibinfo  {journal} {Phys.
  Rev. D}\ }\textbf {\bibinfo {volume} {89}},\ \bibinfo {pages} {084005}
  (\bibinfo {year} {2014})}\BibitemShut {NoStop}%
\bibitem [{\citenamefont {Sennett}\ and\ \citenamefont {Buonanno}(2016)}]{ST4}%
  \BibitemOpen
  \bibfield  {author} {\bibinfo {author} {\bibfnamefont {N.}~\bibnamefont
  {Sennett}}\ and\ \bibinfo {author} {\bibfnamefont {A.}~\bibnamefont
  {Buonanno}},\ }\href {\doibase 10.1103/PhysRevD.93.124004} {\bibfield
  {journal} {\bibinfo  {journal} {Phys. Rev.}\ }\textbf {\bibinfo {volume}
  {D93}},\ \bibinfo {pages} {124004} (\bibinfo {year} {2016})},\ \Eprint
  {http://arxiv.org/abs/1603.03300} {arXiv:1603.03300 [gr-qc]} \BibitemShut
  {NoStop}%
\bibitem [{\citenamefont {Andreou}\ \emph {et~al.}(2019)\citenamefont
  {Andreou}, \citenamefont {Franchini}, \citenamefont {Ventagli},\ and\
  \citenamefont {Sotiriou}}]{Andreou:2019ikc}%
  \BibitemOpen
  \bibfield  {author} {\bibinfo {author} {\bibfnamefont {N.}~\bibnamefont
  {Andreou}}, \bibinfo {author} {\bibfnamefont {N.}~\bibnamefont {Franchini}},
  \bibinfo {author} {\bibfnamefont {G.}~\bibnamefont {Ventagli}}, \ and\
  \bibinfo {author} {\bibfnamefont {T.~P.}\ \bibnamefont {Sotiriou}},\ }\href
  {\doibase 10.1103/PhysRevD.99.124022} {\bibfield  {journal} {\bibinfo
  {journal} {Phys. Rev.}\ }\textbf {\bibinfo {volume} {D99}},\ \bibinfo {pages}
  {124022} (\bibinfo {year} {2019})},\ \Eprint
  {http://arxiv.org/abs/1904.06365} {arXiv:1904.06365 [gr-qc]} \BibitemShut
  {NoStop}%
\bibitem [{\citenamefont {Silva}\ \emph {et~al.}(2019)\citenamefont {Silva},
  \citenamefont {Macedo}, \citenamefont {Sotiriou}, \citenamefont {Gualtieri},
  \citenamefont {Sakstein},\ and\ \citenamefont {Berti}}]{Silva:2018qhn}%
  \BibitemOpen
  \bibfield  {author} {\bibinfo {author} {\bibfnamefont {H.~O.}\ \bibnamefont
  {Silva}}, \bibinfo {author} {\bibfnamefont {C.~F.~B.}\ \bibnamefont
  {Macedo}}, \bibinfo {author} {\bibfnamefont {T.~P.}\ \bibnamefont
  {Sotiriou}}, \bibinfo {author} {\bibfnamefont {L.}~\bibnamefont {Gualtieri}},
  \bibinfo {author} {\bibfnamefont {J.}~\bibnamefont {Sakstein}}, \ and\
  \bibinfo {author} {\bibfnamefont {E.}~\bibnamefont {Berti}},\ }\href
  {\doibase 10.1103/PhysRevD.99.064011} {\bibfield  {journal} {\bibinfo
  {journal} {Phys. Rev.}\ }\textbf {\bibinfo {volume} {D99}},\ \bibinfo {pages}
  {064011} (\bibinfo {year} {2019})},\ \Eprint
  {http://arxiv.org/abs/1812.05590} {arXiv:1812.05590 [gr-qc]} \BibitemShut
  {NoStop}%
\bibitem [{\citenamefont {Macedo}\ \emph {et~al.}(2019)\citenamefont {Macedo},
  \citenamefont {Sakstein}, \citenamefont {Berti}, \citenamefont {Gualtieri},
  \citenamefont {Silva},\ and\ \citenamefont {Sotiriou}}]{Macedo:2019sem}%
  \BibitemOpen
  \bibfield  {author} {\bibinfo {author} {\bibfnamefont {C.~F.~B.}\
  \bibnamefont {Macedo}}, \bibinfo {author} {\bibfnamefont {J.}~\bibnamefont
  {Sakstein}}, \bibinfo {author} {\bibfnamefont {E.}~\bibnamefont {Berti}},
  \bibinfo {author} {\bibfnamefont {L.}~\bibnamefont {Gualtieri}}, \bibinfo
  {author} {\bibfnamefont {H.~O.}\ \bibnamefont {Silva}}, \ and\ \bibinfo
  {author} {\bibfnamefont {T.~P.}\ \bibnamefont {Sotiriou}},\ }\href {\doibase
  10.1103/PhysRevD.99.104041} {\bibfield  {journal} {\bibinfo  {journal} {Phys.
  Rev.}\ }\textbf {\bibinfo {volume} {D99}},\ \bibinfo {pages} {104041}
  (\bibinfo {year} {2019})},\ \Eprint {http://arxiv.org/abs/1903.06784}
  {arXiv:1903.06784 [gr-qc]} \BibitemShut {NoStop}%
\bibitem [{\citenamefont {Blázquez-Salcedo}\ \emph {et~al.}(2018)\citenamefont
  {Blázquez-Salcedo}, \citenamefont {Doneva}, \citenamefont {Kunz},\ and\
  \citenamefont {Yazadjiev}}]{Blazquez-Salcedo:2018jnn}%
  \BibitemOpen
  \bibfield  {author} {\bibinfo {author} {\bibfnamefont {J.~L.}\ \bibnamefont
  {Blázquez-Salcedo}}, \bibinfo {author} {\bibfnamefont {D.~D.}\ \bibnamefont
  {Doneva}}, \bibinfo {author} {\bibfnamefont {J.}~\bibnamefont {Kunz}}, \ and\
  \bibinfo {author} {\bibfnamefont {S.~S.}\ \bibnamefont {Yazadjiev}},\ }\href
  {\doibase 10.1103/PhysRevD.98.084011} {\bibfield  {journal} {\bibinfo
  {journal} {Phys. Rev.}\ }\textbf {\bibinfo {volume} {D98}},\ \bibinfo {pages}
  {084011} (\bibinfo {year} {2018})},\ \Eprint
  {http://arxiv.org/abs/1805.05755} {arXiv:1805.05755 [gr-qc]} \BibitemShut
  {NoStop}%
\bibitem [{\citenamefont {Cunha}\ \emph {et~al.}(2019)\citenamefont {Cunha},
  \citenamefont {Herdeiro},\ and\ \citenamefont {Radu}}]{Cunha:2019dwb}%
  \BibitemOpen
  \bibfield  {author} {\bibinfo {author} {\bibfnamefont {P.~V.~P.}\
  \bibnamefont {Cunha}}, \bibinfo {author} {\bibfnamefont {C.~A.~R.}\
  \bibnamefont {Herdeiro}}, \ and\ \bibinfo {author} {\bibfnamefont
  {E.}~\bibnamefont {Radu}},\ }\href {\doibase 10.1103/PhysRevLett.123.011101}
  {\bibfield  {journal} {\bibinfo  {journal} {Phys. Rev. Lett.}\ }\textbf
  {\bibinfo {volume} {123}},\ \bibinfo {pages} {011101} (\bibinfo {year}
  {2019})},\ \Eprint {http://arxiv.org/abs/1904.09997} {arXiv:1904.09997
  [gr-qc]} \BibitemShut {NoStop}%
\bibitem [{\citenamefont {Collodel}\ \emph {et~al.}(2020)\citenamefont
  {Collodel}, \citenamefont {Kleihaus}, \citenamefont {Kunz},\ and\
  \citenamefont {Berti}}]{Collodel:2019kkx}%
  \BibitemOpen
  \bibfield  {author} {\bibinfo {author} {\bibfnamefont {L.~G.}\ \bibnamefont
  {Collodel}}, \bibinfo {author} {\bibfnamefont {B.}~\bibnamefont {Kleihaus}},
  \bibinfo {author} {\bibfnamefont {J.}~\bibnamefont {Kunz}}, \ and\ \bibinfo
  {author} {\bibfnamefont {E.}~\bibnamefont {Berti}},\ }\href {\doibase
  10.1088/1361-6382/ab74f9} {\bibfield  {journal} {\bibinfo  {journal} {Class.
  Quant. Grav.}\ }\textbf {\bibinfo {volume} {37}},\ \bibinfo {pages} {075018}
  (\bibinfo {year} {2020})},\ \Eprint {http://arxiv.org/abs/1912.05382}
  {arXiv:1912.05382 [gr-qc]} \BibitemShut {NoStop}%
\bibitem [{\citenamefont {Damour}\ \emph {et~al.}(1976)\citenamefont {Damour},
  \citenamefont {Deruelle},\ and\ \citenamefont {Ruffini}}]{Damour:1976}%
  \BibitemOpen
  \bibfield  {author} {\bibinfo {author} {\bibfnamefont {T.}~\bibnamefont
  {Damour}}, \bibinfo {author} {\bibfnamefont {N.}~\bibnamefont {Deruelle}}, \
  and\ \bibinfo {author} {\bibfnamefont {R.}~\bibnamefont {Ruffini}},\ }\href
  {\doibase 10.1007/BF02725534} {\bibfield  {journal} {\bibinfo  {journal}
  {Lettere al Nuovo Cimento}\ }\textbf {\bibinfo {volume} {15}},\ \bibinfo
  {pages} {257} (\bibinfo {year} {1976})}\BibitemShut {NoStop}%
\bibitem [{\citenamefont {Zouros}\ and\ \citenamefont
  {Eardley}(1979)}]{Zouros:1979iw}%
  \BibitemOpen
  \bibfield  {author} {\bibinfo {author} {\bibfnamefont {T.~J.~M.}\
  \bibnamefont {Zouros}}\ and\ \bibinfo {author} {\bibfnamefont {D.~M.}\
  \bibnamefont {Eardley}},\ }\href {\doibase 10.1016/0003-4916(79)90237-9}
  {\bibfield  {journal} {\bibinfo  {journal} {Annals Phys.}\ }\textbf {\bibinfo
  {volume} {118}},\ \bibinfo {pages} {139} (\bibinfo {year}
  {1979})}\BibitemShut {NoStop}%
\bibitem [{\citenamefont {Detweiler}(1980)}]{Detweiler:1980uk}%
  \BibitemOpen
  \bibfield  {author} {\bibinfo {author} {\bibfnamefont {S.~L.}\ \bibnamefont
  {Detweiler}},\ }\href {\doibase 10.1103/PhysRevD.22.2323} {\bibfield
  {journal} {\bibinfo  {journal} {Phys. Rev.}\ }\textbf {\bibinfo {volume}
  {D22}},\ \bibinfo {pages} {2323} (\bibinfo {year} {1980})}\BibitemShut
  {NoStop}%
\bibitem [{\citenamefont {Dolan}(2007)}]{Dolan:2007mj}%
  \BibitemOpen
  \bibfield  {author} {\bibinfo {author} {\bibfnamefont {S.~R.}\ \bibnamefont
  {Dolan}},\ }\href {\doibase 10.1103/PhysRevD.76.084001} {\bibfield  {journal}
  {\bibinfo  {journal} {Phys. Rev.}\ }\textbf {\bibinfo {volume} {D76}},\
  \bibinfo {pages} {084001} (\bibinfo {year} {2007})},\ \Eprint
  {http://arxiv.org/abs/0705.2880} {arXiv:0705.2880 [gr-qc]} \BibitemShut
  {NoStop}%
\bibitem [{\citenamefont {Dima}\ and\ \citenamefont
  {Barausse}(2020)}]{Dima:2020rzg}%
  \BibitemOpen
  \bibfield  {author} {\bibinfo {author} {\bibfnamefont {A.}~\bibnamefont
  {Dima}}\ and\ \bibinfo {author} {\bibfnamefont {E.}~\bibnamefont
  {Barausse}},\ }\href@noop {} {\  (\bibinfo {year} {2020})},\ \Eprint
  {http://arxiv.org/abs/2001.11484} {arXiv:2001.11484 [gr-qc]} \BibitemShut
  {NoStop}%
\bibitem [{\citenamefont {Dolan}(2013)}]{Dolan:2012yt}%
  \BibitemOpen
  \bibfield  {author} {\bibinfo {author} {\bibfnamefont {S.~R.}\ \bibnamefont
  {Dolan}},\ }\href {\doibase 10.1103/PhysRevD.87.124026} {\bibfield  {journal}
  {\bibinfo  {journal} {Phys. Rev.}\ }\textbf {\bibinfo {volume} {D87}},\
  \bibinfo {pages} {124026} (\bibinfo {year} {2013})},\ \Eprint
  {http://arxiv.org/abs/1212.1477} {arXiv:1212.1477 [gr-qc]} \BibitemShut
  {NoStop}%
\bibitem [{\citenamefont {Varshalovich}\ \emph {et~al.}(1988)\citenamefont
  {Varshalovich}, \citenamefont {Moskalev},\ and\ \citenamefont
  {Khersonskii}}]{AngMom}%
  \BibitemOpen
  \bibfield  {author} {\bibinfo {author} {\bibfnamefont {D.~A.}\ \bibnamefont
  {Varshalovich}}, \bibinfo {author} {\bibfnamefont {A.~N.}\ \bibnamefont
  {Moskalev}}, \ and\ \bibinfo {author} {\bibfnamefont {V.~K.}\ \bibnamefont
  {Khersonskii}},\ }\href {\doibase 10.1142/0270} {\emph {\bibinfo {title}
  {Quantum Theory of Angular Momentum}}}\ (\bibinfo  {publisher} {WORLD
  SCIENTIFIC},\ \bibinfo {year} {1988})\BibitemShut {NoStop}%
\bibitem [{\citenamefont {Palenzuela}\ \emph {et~al.}(2009)\citenamefont
  {Palenzuela}, \citenamefont {Lehner}, \citenamefont {Reula},\ and\
  \citenamefont {Rezzolla}}]{Palenzuela:2008sf}%
  \BibitemOpen
  \bibfield  {author} {\bibinfo {author} {\bibfnamefont {C.}~\bibnamefont
  {Palenzuela}}, \bibinfo {author} {\bibfnamefont {L.}~\bibnamefont {Lehner}},
  \bibinfo {author} {\bibfnamefont {O.}~\bibnamefont {Reula}}, \ and\ \bibinfo
  {author} {\bibfnamefont {L.}~\bibnamefont {Rezzolla}},\ }\href {\doibase
  10.1111/j.1365-2966.2009.14454.x} {\bibfield  {journal} {\bibinfo  {journal}
  {Mon. Not. Roy. Astron. Soc.}\ }\textbf {\bibinfo {volume} {394}},\ \bibinfo
  {pages} {1727} (\bibinfo {year} {2009})},\ \Eprint
  {http://arxiv.org/abs/0810.1838} {arXiv:0810.1838 [astro-ph]} \BibitemShut
  {NoStop}%
\bibitem [{\citenamefont {Pareschi}\ and\ \citenamefont
  {Russo}(2005)}]{PareschiRusso}%
  \BibitemOpen
  \bibfield  {author} {\bibinfo {author} {\bibfnamefont {L.}~\bibnamefont
  {Pareschi}}\ and\ \bibinfo {author} {\bibfnamefont {G.}~\bibnamefont
  {Russo}},\ }\href {\doibase 10.1007/s10915-004-4636-4} {\bibfield  {journal}
  {\bibinfo  {journal} {Journal of Scientific Computing}\ }\textbf {\bibinfo
  {volume} {25}},\ \bibinfo {pages} {129} (\bibinfo {year} {2005})}\BibitemShut
  {NoStop}%
\bibitem [{\citenamefont {Berti}\ \emph {et~al.}(2009)\citenamefont {Berti},
  \citenamefont {Cardoso},\ and\ \citenamefont {Starinets}}]{Berti:2009kk}%
  \BibitemOpen
  \bibfield  {author} {\bibinfo {author} {\bibfnamefont {E.}~\bibnamefont
  {Berti}}, \bibinfo {author} {\bibfnamefont {V.}~\bibnamefont {Cardoso}}, \
  and\ \bibinfo {author} {\bibfnamefont {A.~O.}\ \bibnamefont {Starinets}},\
  }\href {\doibase 10.1088/0264-9381/26/16/163001} {\bibfield  {journal}
  {\bibinfo  {journal} {Class. Quant. Grav.}\ }\textbf {\bibinfo {volume}
  {26}},\ \bibinfo {pages} {163001} (\bibinfo {year} {2009})},\ \Eprint
  {http://arxiv.org/abs/0905.2975} {arXiv:0905.2975 [gr-qc]} \BibitemShut
  {NoStop}%
\bibitem [{\citenamefont {Herdeiro}\ \emph {et~al.}(2020)\citenamefont
  {Herdeiro}, \citenamefont {Radu}, \citenamefont {Silva}, \citenamefont
  {Sotiriou},\ and\ \citenamefont {Yunes}}]{Herdeiro:2020wei}%
  \BibitemOpen
  \bibfield  {author} {\bibinfo {author} {\bibfnamefont {C.~A.}\ \bibnamefont
  {Herdeiro}}, \bibinfo {author} {\bibfnamefont {E.}~\bibnamefont {Radu}},
  \bibinfo {author} {\bibfnamefont {H.~O.}\ \bibnamefont {Silva}}, \bibinfo
  {author} {\bibfnamefont {T.~P.}\ \bibnamefont {Sotiriou}}, \ and\ \bibinfo
  {author} {\bibfnamefont {N.}~\bibnamefont {Yunes}},\ }\href@noop {} {\
  (\bibinfo {year} {2020})},\ \Eprint {http://arxiv.org/abs/2009.03904}
  {arXiv:2009.03904 [gr-qc]} \BibitemShut {NoStop}%
\bibitem [{\citenamefont {Berti}\ \emph {et~al.}(2020)\citenamefont {Berti},
  \citenamefont {Collodel}, \citenamefont {Kleihaus},\ and\ \citenamefont
  {Kunz}}]{Berti:2020kgk}%
  \BibitemOpen
  \bibfield  {author} {\bibinfo {author} {\bibfnamefont {E.}~\bibnamefont
  {Berti}}, \bibinfo {author} {\bibfnamefont {L.~G.}\ \bibnamefont {Collodel}},
  \bibinfo {author} {\bibfnamefont {B.}~\bibnamefont {Kleihaus}}, \ and\
  \bibinfo {author} {\bibfnamefont {J.}~\bibnamefont {Kunz}},\ }\href@noop {}
  {\  (\bibinfo {year} {2020})},\ \Eprint {http://arxiv.org/abs/2009.03905}
  {arXiv:2009.03905 [gr-qc]} \BibitemShut {NoStop}%
\bibitem [{\citenamefont {Bambi}\ and\ \citenamefont
  {Barausse}(2011)}]{Bambi:2011jq}%
  \BibitemOpen
  \bibfield  {author} {\bibinfo {author} {\bibfnamefont {C.}~\bibnamefont
  {Bambi}}\ and\ \bibinfo {author} {\bibfnamefont {E.}~\bibnamefont
  {Barausse}},\ }\href {\doibase 10.1088/0004-637X/731/2/121} {\bibfield
  {journal} {\bibinfo  {journal} {Astrophys. J.}\ }\textbf {\bibinfo {volume}
  {731}},\ \bibinfo {pages} {121} (\bibinfo {year} {2011})},\ \Eprint
  {http://arxiv.org/abs/1012.2007} {arXiv:1012.2007 [gr-qc]} \BibitemShut
  {NoStop}%
\bibitem [{\citenamefont {Ventagli}\ \emph {et~al.}(2020)\citenamefont
  {Ventagli}, \citenamefont {Lehebel},\ and\ \citenamefont
  {Sotiriou}}]{Ventagli:2020rnx}%
  \BibitemOpen
  \bibfield  {author} {\bibinfo {author} {\bibfnamefont {G.}~\bibnamefont
  {Ventagli}}, \bibinfo {author} {\bibfnamefont {A.}~\bibnamefont {Lehebel}}, \
  and\ \bibinfo {author} {\bibfnamefont {T.~P.}\ \bibnamefont {Sotiriou}},\
  }\href@noop {} {\  (\bibinfo {year} {2020})},\ \Eprint
  {http://arxiv.org/abs/2006.01153} {arXiv:2006.01153 [gr-qc]} \BibitemShut
  {NoStop}%
\bibitem [{\citenamefont {Antoniou}\ \emph {et~al.}(2020)\citenamefont
  {Antoniou}, \citenamefont {Bordin},\ and\ \citenamefont
  {Sotiriou}}]{Antoniou:2020nax}%
  \BibitemOpen
  \bibfield  {author} {\bibinfo {author} {\bibfnamefont {G.}~\bibnamefont
  {Antoniou}}, \bibinfo {author} {\bibfnamefont {L.}~\bibnamefont {Bordin}}, \
  and\ \bibinfo {author} {\bibfnamefont {T.~P.}\ \bibnamefont {Sotiriou}},\
  }\href@noop {} {\  (\bibinfo {year} {2020})},\ \Eprint
  {http://arxiv.org/abs/2004.14985} {arXiv:2004.14985 [gr-qc]} \BibitemShut
  {NoStop}%
\end{thebibliography}%

\end{document}